\documentclass[conference]{IEEEtran}
\IEEEoverridecommandlockouts
\usepackage{cite}
\usepackage{amsmath,amssymb,amsfonts}
\usepackage{algorithmic}
\usepackage{graphicx}
\usepackage{textcomp}
\usepackage{xcolor}
\usepackage{multirow}
\usepackage{amsmath}
\usepackage{subcaption}
\usepackage{fixltx2e}
\usepackage{float}
\usepackage{siunitx}
\usepackage{soul}
\usepackage{colortbl}
\usepackage{authblk}
\usepackage{makecell}
\usepackage{gensymb}
\usepackage[shortlabels]{enumitem}

\usepackage[bookmarks=false]{hyperref}
\hypersetup{colorlinks,urlcolor=blue}
\hypersetup{nolinks=true}

\usepackage{cleveref}

\def\BibTeX{{\rm B\kern-.05em{\sc i\kern-.025em b}\kern-.08em
    T\kern-.1667em\lower.7ex\hbox{E}\kern-.125emX}}
\begin{document}

\title{Prediction of inter packet arrival times for enhanced NR-V2X sidelink scheduling}

\author{Brian McCarthy}
\author{Aisling O'Driscoll}

\affil{{\textit{School of Computer Science and Information Technology}, \textit{University College Cork}, Cork, Ireland}\\\{b.mccarthy, a.odriscoll\}@cs.ucc.ie}

\maketitle
\begin{abstract}
A significant limitation of the LTE-V2X and NR-V2X sidelink scheduling mechanisms is their difficulty coping with variations in inter packet arrival times, also known as aperiodic packets. This conflicts with the fundamental characteristics of most V2X services which are triggered based on an event. e.g. ETSI Cooperative Awareness Messages (CAMs) - vehicle kinematics, Cooperative Perception Messages (CPMs) - object sensing and Decentralised Event Notification Messages (DENMs) - event occurrences. Furthermore, network management techniques such as congestion control mechanisms can result in varied inter packet arrival times. 
To combat this, NR-V2X introduced a dynamic grant mechanism, which we show is ineffective unless there is background periodic traffic to stabilise the sensing history upon which the scheduler makes it decisions. The characteristics of V2X services make it implausible that such periodic application traffic will exist. 

To overcome this significant drawback, we demonstrate that the standardised scheduling algorithms can be made effective if the event triggered arrival rate of packets can be accurately predicted. These predictions can be used to tune the Resource Reservation Interval (RRI) parameter of the MAC scheduler to negate the negative impact of aperiodicity. Such an approach allows the scheduler to achieve comparable performance to a scenario where packets arrive periodically. To demonstrate the effectiveness of our approach, an ML model has been devised for the prediction of cooperative awareness messages, but the same principle can be abstracted to other V2X service types.
\end{abstract} 

\begin{IEEEkeywords}
Cellular V2X (C-V2X), New Radio (NR-V2X), sensing based semi persistent scheduling (SB-SPS), dynamic grant, aperiodic arrival rates, sidelink.

\end{IEEEkeywords}

\section{Introduction}
\label{sec:intro}
The latest cellular vehicular communication standard, NR-V2X Release 16 \cite{3gpp-rel16}, is an evolution of the earlier Cellular V2X/LTE-V2X (C-V2X) standards specified in the Third Generation Partnership Project (3GPP) Releases 14 \cite{3gpp} and 15 \cite{3gpp-rel15}. It is envisaged that these standards will deliver improvements in traffic safety and efficiency for connected vehicles through direct communication (V2V) and communication with infrastructure (V2I). They are also designed to facilitate the shift towards increased autonomy, supporting enhanced V2X (e-V2X) services such as cooperative perception, vehicle platooning and remote driving, amongst others \cite{3gpp-rel16-apps}. While NR-V2X introduces several significant improvements to C-V2X to support higher levels of autonomy, it does not significantly change the underlying MAC sidelink scheduling algorithm, Sensing-Based Semi-Persistent Scheduling (SB-SPS). This means that NR-V2X inherits the drawbacks of the underlying scheduling approach. One significant drawback of SB-SPS lies in its design, which assumes the periodic arrival of application packets. Accurate estimation of which radio resources might be free in the future depends on the correct recording of a sensing history within the scheduler so that grants i.e. reserved resources, can be maintained. It has been shown by the authors of this paper \cite{our-vehits} and others in the vehicular communications community \cite{molina-aperiodic,bazzi-aperiodic,icc-repetition} that packets exhibiting aperiodic arrival rates lead to inaccuracies in the sensing history, causing under-utilisation of radio resources and instability in scheduling.  

This issue is significant as the application characteristics of vehicular services are inherently aperiodic in real world scenarios. One of the most fundamental V2X services is based on the exchange of ETSI specified Cooperative Awareness Messages (CAMs). This forms the basis for vehicular safety applications, where vehicles communicate their current dynamics with each other in order to avoid collisions. It is commonly assumed that CAMs are sent periodically, typically every 100ms. However this has been widely debunked over the last decade in the vehicular communications community, who have proven that periodic transmissions can lead to a broadcast storm, especially when vehicle density increases \cite{analysis-old-beaconing, sommar-beaconing}. In recognition of this, ETSI has specified a set of CAM trigger conditions based on the vehicles kinematics to suppress unnecessary transmissions. These triggering rules result in the application packets arriving in an aperiodic pattern. As NR-V2X SB-SPS cannot effectively schedule these aperiodic packets, this introduces an incompatibility with current CA services. A similar scenario exists for services such as Cooperative Perception (CP) which are based on sensed object dynamics \cite{etsi-cpm}, as well as DENMs which are event triggered. Congestion control mechanisms, necessary to limit network load given the scarce dedicated spectrum in the ITS bands, also lead to aperiodic arrival rates \cite{our-wowmom}. 

To combat these limitations, the 3GPP have introduced some significant changes to the SB-SPS scheduler, such as dynamic grant. However we show that this approach does not solve the issue unless some background packets arrive according to a periodic rate; a scenario that cannot be guaranteed and is unlikely. This paper suggests an alternative approach that allows the scheduler to predict packet arrival rates based on the vehicle dynamics and thus schedule radio resources as if packets were arriving in a pre-determined periodic fashion. It does this by dynamically maintaining the SB-SPS resource reservation interval (RRI), a parameter that is used by SB-SPS to indicate the period between packet transmissions. The means by which this is maintained is via a deep-learning recurrent neural network. The result is that SB-SPS delivers comparable performance for packets arriving according to an aperiodic inter-packet arrival rate to periodic transmissions, negating the unintended negative scheduling decisions inherent to the SB-SPS grant mechanism.

There have been two papers that have investigated this issue and proposed solutions based on machine learning approaches. Lusvarghi et. al. \cite{lusvarghi-prediction} present the first work in this area with a k-Nearest Neighbours (KNN) model that incorporates a change to the standardised SB-SPS process. The authors propose that the SB-SPS reselection counter (\textit{C\textsubscript{resel}}) be included in the SCI. This additional parameter specifies the remaining transmissions in the grant i.e. indicating how long a vehicle will maintain its grant. In the grant selection process, \textit{C\textsubscript{resel}} is calculated by predicting how long a sequence of the same RRI will be maintained. This is potentially problematic as variable vehicle dynamics within a single grant may result in a highly variable sequence of RRIs resulting in short-lived grants. In contrast the proposed model can handle such cases more easily. The author's solution also relies on accurate knowledge of the speed and position of the preceding vehicle. While comprehensively evaluated, it is not clear whether the proposed model would generalise for unseen scenarios as only a highway scenario is investigated outside of their original dataset. Finally, this model would require retraining for any CAM rule change or for rule changes associated with different services, whereas the proposed model predictions would be applicable to any rule set, due to it's reliance only on mobility prediction and not the specific CAM time trigger.

The second solution is proposed by Seon et. al. \cite{Seon-prediction}. They propose a Multilayer Perceptron (MLP) model, which is a form of neural network commonly used in image classification problems, but also applicable for time series data. The model directly predicts the CAM time by incorporating speed, position, heading and time changes, as well as the previous CAM trigger condition. The model is trained on two real world vehicle mobility data sets representing mobility traces of urban and suburban driving scenarios. Three vehicle traces are in the urban scenario and the remaining 7 traces are suburban. As such, this represents a relatively small dataset. The model shows high levels of accuracy in all cases across trained, validation and test datasets. A drawback of this study is that the results presented relating to the improvement of SB-SPS are reported on the test set. The test set is formed from the original traces and as such is likely to be similar to the training data. This therefore cannot show the effectiveness of the model in generalising to unseen data. As such the model may not prove to be effective outside of these or similar scenarios. 

The rest of this paper is organised as follows; Section \ref{sec:applicationBackground} introduces the negative impact of the application characteristics on the NR-V2X and C-V2X scheduling mechanisms. Section \ref{sec:machineLearning} introduces the machine learning approach adopted GRU and the model used in the study. Section \ref{sec:results} includes experimental results adopting this proposed solution. Section \ref{sec:conclusion} summarises the paper conclusions.

\begin{table}[htbp]
\caption{V2X Abbreviations and Acronyms.}
\label{acronymtable}
\begin{center}
\renewcommand{\arraystretch}{1.2}
\begin{tabular}{l | p{6cm}}
\hline\hline 
\textbf{Acronym} & \textbf{Description} \\ [0.5ex] 
\hline
\multicolumn{2}{c}{\textbf{Standards}}\\
\hline
LTE-V/C-V2X & 3GPP Release 14 standard for vehicular communications (based on LTE).\\
\hline
NR-V2X & 3GPP Release 16 standard for vehicular communications. \\
\hline
SB-SPS & Sensing Based Semi-Persistent Scheduling (uses a grant mechanism).\\
\hline
\multicolumn{2}{c}{\textbf{Scheduling}}\\
\hline
SCI & Sidelink Control Information.\\
\hline
CSR & Candidate single Subframe Resource. Can be one or more subchannel.\\
\hline
RRI & Resource Reservation Interval. Period between transmissions.\\
\hline
GB / No GB & SB-SPS Grant Breaking parameter. Defines whether a subchannel is maintained or not.\\
\hline
\multicolumn{2}{c}{\textbf{Services}}\\
\hline
CAM & Cooperative Awareness Message - Triggered on Vehicle Kinematics\\
\hline
CPM & Cooperative Perception Message - Triggered based on sensed object kinematics\\
\hline
DENM & Decentralized Environmental Notification - Event triggered message.\\
\hline
\multicolumn{2}{c}{\textbf{Metrics}}\\
\hline
PDR & Packet Delivery Rate.\\
\hline
IPT & Inter-Packet Time, time between transmissions of a packet.\\
\hline\hline
\end{tabular}
\end{center}
\end{table} 

\section{The impact of aperiodic inter-packet arrival times on the C-V2X/NR-V2X scheduler}
\label{sec:applicationBackground}



To quantitatively contextualise the need for the proposed solution, we simulate the performance of the SB-SPS algorithm with the ETSI CAM service to highlight the limitations of the SB-SPS scheduler. 

\begin{figure}[htbp]
\begin{subfigure}{.48\textwidth}
  \centering
  \includegraphics[width=.88\linewidth]{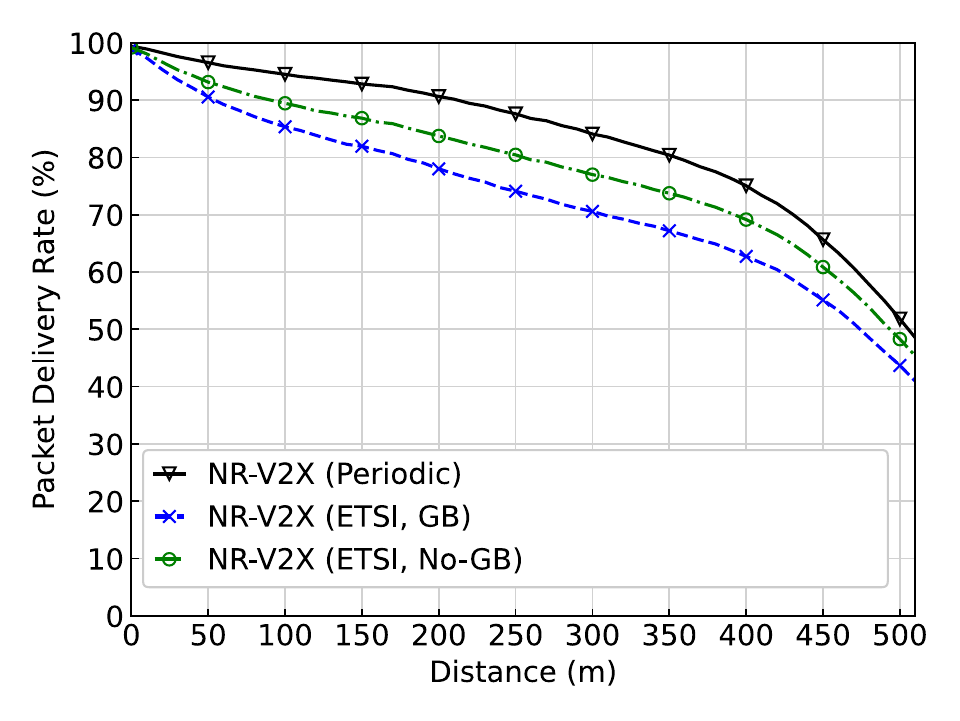}
  \caption{NR-V2X SB-SPS.}
  \label{fig:nr-v2x}
\end{subfigure}
\begin{subfigure}{.48\textwidth}
  \centering
  \includegraphics[width=.95\linewidth]{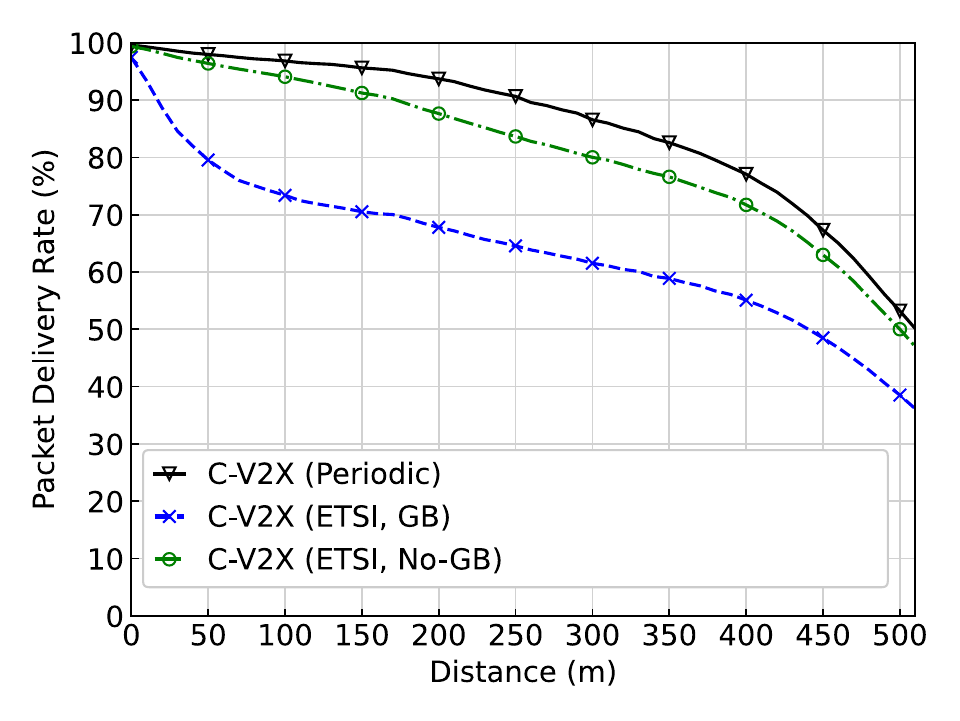}
  \caption{C-V2X SB-SPS.}
  \label{fig:c-v2x}
\end{subfigure}
\caption{Impact of aperiodic inter-packet times (IPT) on Cellular Vehicular Standards.}
\label{fig:Aperiodic-Performance}
\end{figure}

Fig. \ref{fig:Aperiodic-Performance} highlights the performance issues for both C-V2X and NR-V2X when dealing with this aperiodic traffic pattern. A substantial decline in PDR is noted for C-V2X and to a lesser extent for NR-V2X. The cause of the performance degradation is as a result of missed transmissions i.e. when there is no packet to send but the granted resource becomes available. This results in a grant break (labelled GB), which causes additional contention and increased collisions. The alternative approach is to disable grant breaking (labelled No-GB), which improves performance but also results in collisions due to an inconsistent sensing history utilised in the selection process.

\begin{figure}
    \centering
    \includegraphics[width=\linewidth]{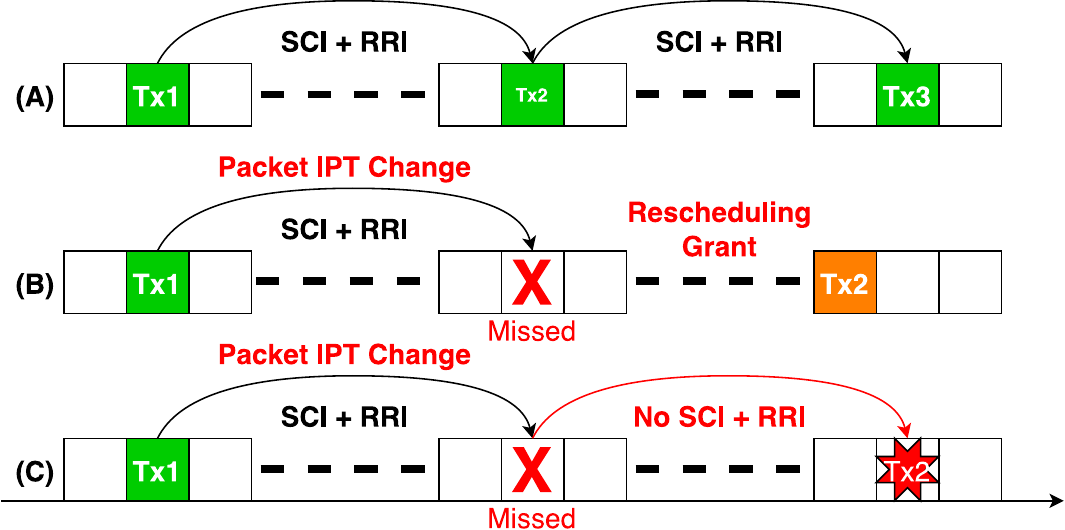}
    \caption{SB-SPS Scheduling with a) Periodic inter-packet times (IPT) B) Aperiodic IPT with grant breaking, and C) Aperiodic IPT with no grant breaking.}
    \label{fig:sb-sps-explanation}
\end{figure}

The cause for these performance degradations is illustrated in Fig. \ref{fig:sb-sps-explanation}. Timeline A represents the default performance of SB-SPS with periodic traffic. The SB-SPS grant mechanism is maintained by a vehicle transmitting a SCI message that includes a RRI field. The RRI reserves a resource (subchannel(s)) for the next time period. On the termination of the grant the RRI field will be 0, relinquishing the subchannel(s). This RRI field is vital to the performance of SB-SPS, as it indicates to other vehicles  which resources are reserved in the future and thus are to be avoided when scheduling.

Issues arise with the introduction of aperiodic traffic patterns e.g. CAMs. Timeline \textit{B} indicates SB-SPS performance with GB enabled. An SCI is sent at Tx1 reserving a subchannel in a later time slot. However, due to vehicle dynamics, a packet is not available to transmit causing a missed transmission opportunity. The grant is relinquished after the missed transmission opportunity and a new grant will be established upon the arrival of the next application packet (marked as Tx2 in orange). This increases channel contention and ultimately collisions \cite{he-aperiodic-analytical}. 

Timeline \textit{C} does not grant break (No-GB) and as such the grant is maintained through the missed transmission slot. While this does not increase contention, the lack of an SCI message and the associated RRI, results in neighbouring vehicles viewing the resource as being available. Thus if a packet is transmitted at Tx2, a collision may occur with a neighbouring vehicle (indicated in red). Additionally in both timelines, missed transmission opportunities (indicated by the red X) represent a reserved resource which is unutilised and as such represent wasted channel resources.

\begin{figure}
  \centering
  \includegraphics[width=.95\linewidth]{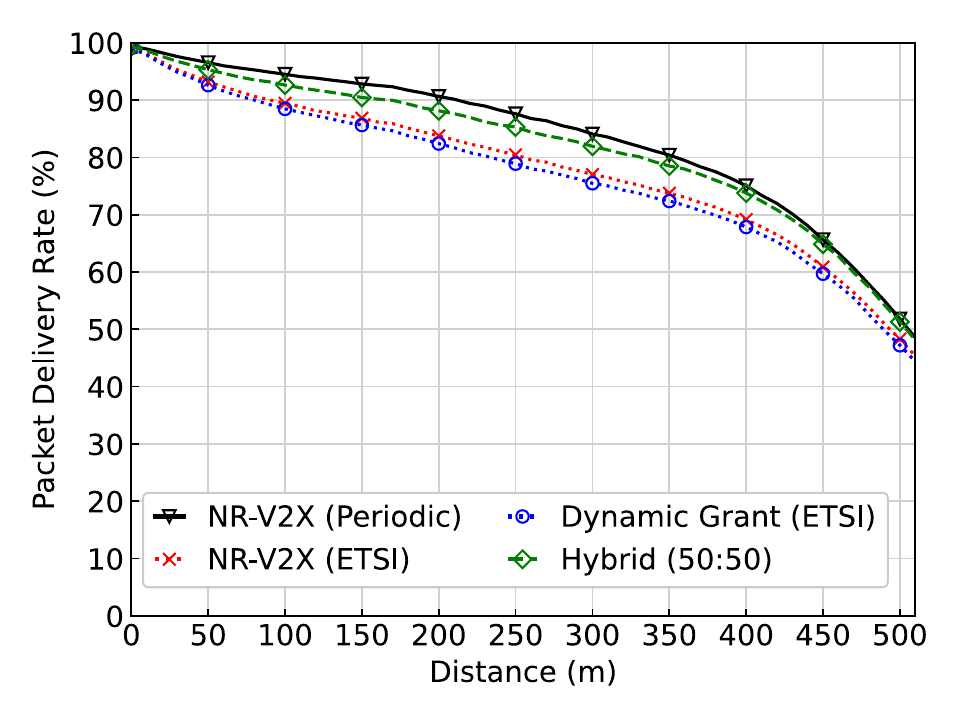}
  \caption{NR-V2X Dynamic Grant Performance for periodic, aperiodic and hybrid application traffic.}
  \label{fig:Dynamic-Hybrid}
\end{figure}

NR-V2X introduces 2 novelties to the C-V2X standard; Resource pre-emption and Dynamic Grant. Resource pre-emption occurs when a vehicle senses that one of its reserved resources is likely to be needed by another vehicle's higher priority application traffic. In such cases, the resource can be relinquished. The vehicle must then engage in resource rescheduling for its lower priority traffic. This mechanism is designed to offer greater flexibility when dealing with traffic of mixed of priority e.g. DENMs which have a higher priority than CAMs or CPMs which have similar priorities. Dynamic Grant is designed to address aperiodic arrival rates. It uses the same underlying SB-SPS mechanism but only reserves a single resource, rather than maintaining a persistent grant. We show in Fig. \ref{fig:Dynamic-Hybrid} that this is not effective unless high levels of background periodic traffic exists upon which to base the scheduling decisions. Fig. \ref{fig:Dynamic-Hybrid} considers fully aperiodic traffic and a hybrid of periodic and aperiodic packets (50\%:50\%). It is evident that dynamic grant can perform well with high levels of periodic background traffic, i.e. 50\% periodic traffic, but cannot resolve issues with exclusively aperiodic application traffic i.e. ETSI.



Ultimately we conclude that dynamic grant cannot inherently manage high levels of aperiodic traffic. A solution to manage equal priority services is vital as it will account for the majority of application traffic for initial vehicular deployments and the services underpinning increased autonomy.

\section{A Deep learning model for CA IPT prediction and enhanced scheduling}
\label{sec:machineLearning}

The objective of our proposed approach is to enable the accurate prediction of aperiodic inter-packet times (IPT). This prediction can be used to overcome the difficulties just outlined.

\subsection{Enhanced scheduling via CA IPT Prediction}
\label{subsec:solutionOutline}

\begin{figure}
    \centering
    \includegraphics[width=\linewidth]{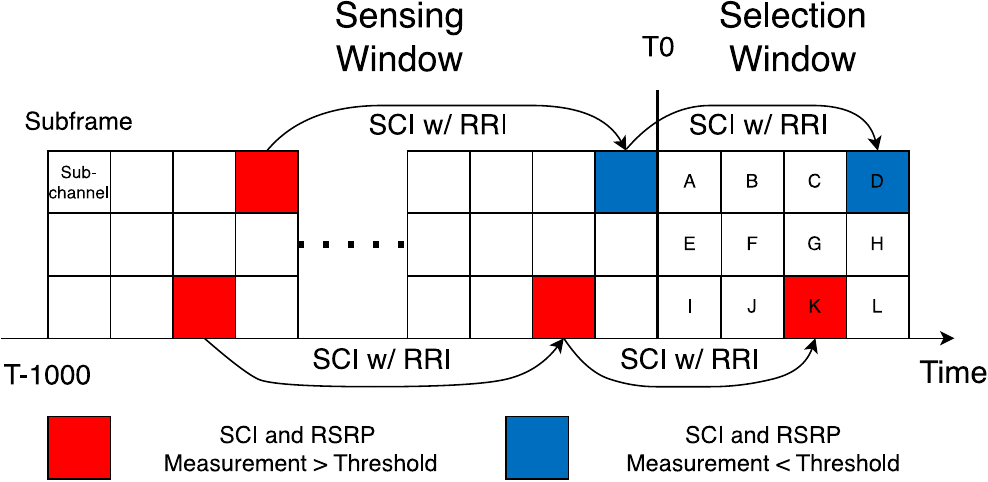}
    \caption{NR-V2X Default SB-SPS Scheduler.}
    \label{fig:sb-sps-mech}
\end{figure}

This section describes how CA prediction can be used to help the SB-SPS scheduler negate the impact of aperiodic arrival rates. The behaviour of the default SB-SPS scheduler is illustrated in Fig. \ref{fig:sb-sps-mech}. SB-SPS uses a grant based approach to scheduling, where each vehicle reserves a resource (a subchannel or set of subchannels in a single subframe) for a number of transmission opportunities. The period between these transmissions is advertised to others in the RRI field of the SCI packet as previously outlined. Upon grant generation, each vehicle uses two windows to determine availability of resources. The first is the sensing window, which is a historical window of all reserved subframes and subchannels that records both the SCI information received as well as the channel conditions i.e. RSRP and RSSI measurements. The second window is a selection window which represents the possible resources that can be selected in the future. This is filtered based on the sensing window. In NR-V2X, the process is different to C-V2X. A vehicle will check the used resources in the sensing window to see if an SCI with RRI has been sent. This determines if a resource might be reserved in the future selection window (red and blue resources in Fig. \ref{fig:sb-sps-mech}). Once the resources that might be reserved are determined, SB-SPS checks if the recorded RSRP exceeds a pre-configured threshold. If so, a resource is discarded from consideration (represented by the K resource in red). Alternatively the resource can potentially be chosen if it does not exceed the threshold (D resource in blue). Ultimately, all resources that are not removed will be reported to the MAC layer where a selection is made at random. As already discussed, the RRI is a vital part of this process as it is the means by which the vehicles are able to determine which resources in the sensing window may be reserved by other vehicles. As such, when the RRI becomes inaccurate this has a significant impact on the performance of the scheduler.

\begin{figure}
    \centering
    \includegraphics[width=\linewidth]{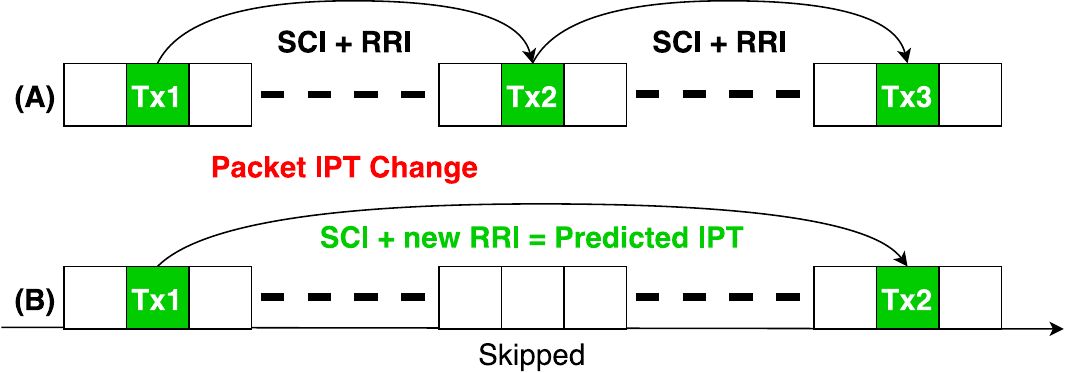}
    \caption{NR-V2X with (A) Default RRI and (B) Predicted RRI.}
    \label{fig:prediction-Solution}
\end{figure}

To overcome the negative impact of an incorrect RRI on the scheduling process, this paper proposes the use of a predictive model which will enable the dynamic update of the RRI to correspond with the IPT of the application. The underlying tenet of this proposal is illustrated in Fig. \ref{fig:prediction-Solution}. Standard SB-SPS is shown in timeline A, where each resource is reserved in a periodic fashion with a corresponding RRI. Timeline B then represents the proposed solution where the packet IPT is predicted upon each packet transmission with the RRI field updated accordingly. Assuming the predictions prove to be highly accurate, this returns SB-SPS to a state similar to that of timeline A.

\subsection{Proposed Gated Recurrent Unit Model (GRU) Model for Prediction of CA service}
\label{subsec:machineLearningBackground}

\begin{figure}[htbp]
\begin{subfigure}{.48\textwidth}
  \centering
  \includegraphics[width=.88\linewidth]{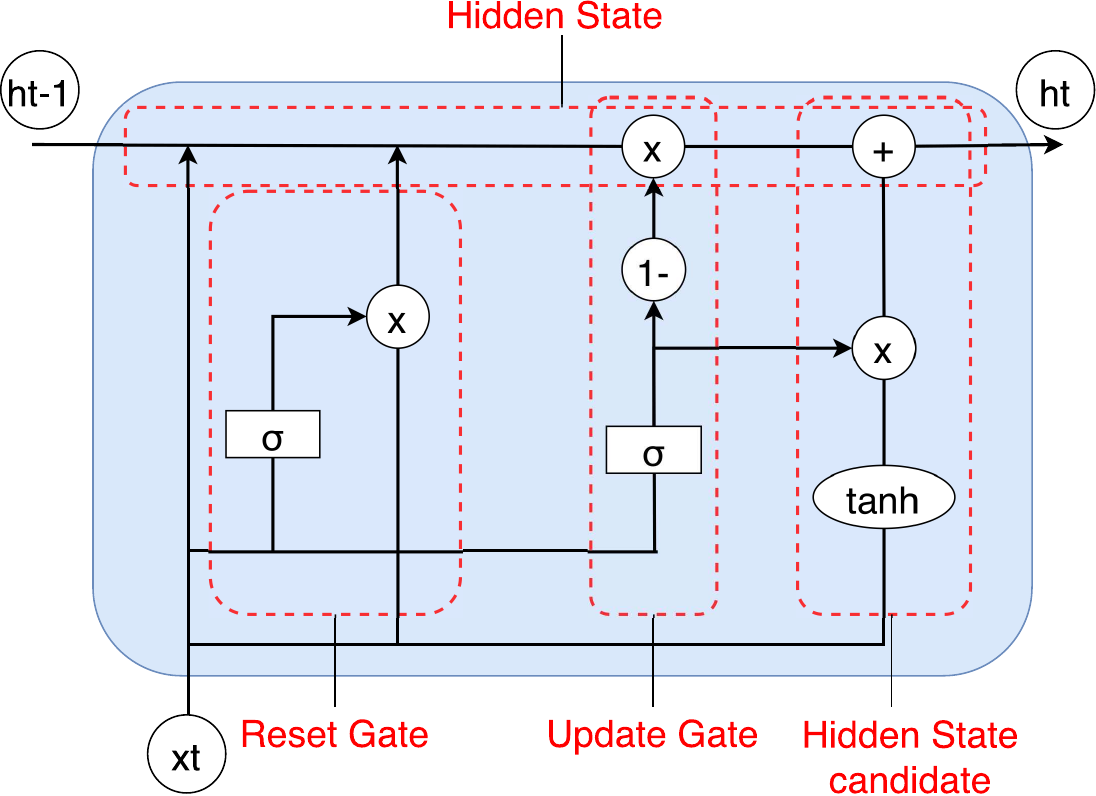}
  \caption{GRU single neuron Architecture.}
  \label{fig:GRU-Arch}
\end{subfigure}
\begin{subfigure}{.48\textwidth}
  \centering
  \includegraphics[width=.95\linewidth]{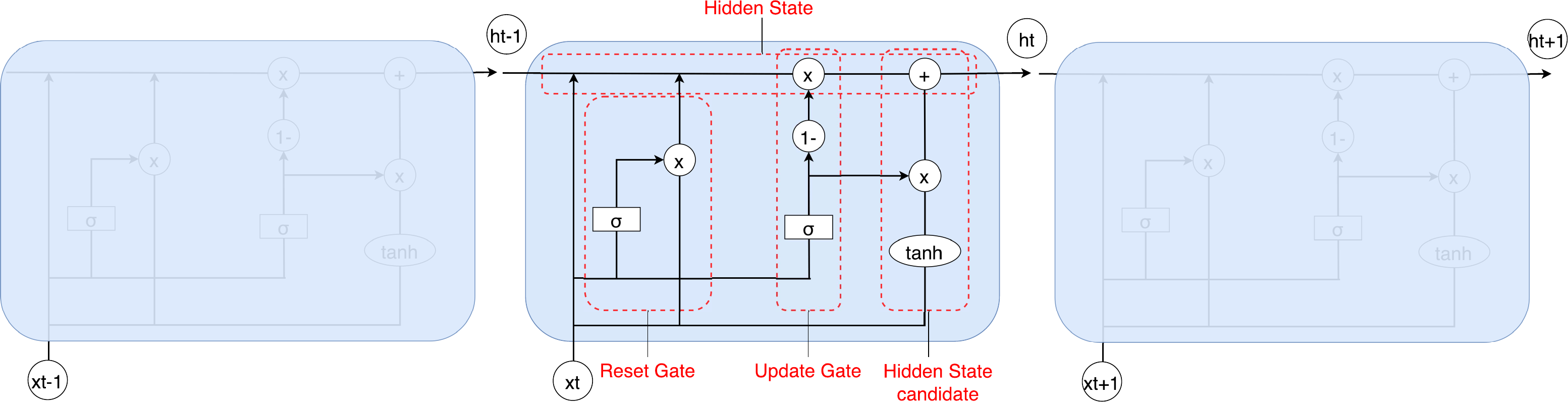}
  \caption{GRU neuron connections.}
  \label{fig:GRU-rolled}
\end{subfigure}
\caption{GRU (RNN) Architecture diagram.}
\label{fig:LSTM-diagram}
\end{figure}

Having outlined how the scheduler will use the CA IPT prediction, this section now describes how the prediction occurs. 
Several studies have looked at the use of Long-Short Term Memory (LSTM) for modelling mobility and trajectory prediction \cite{wang-trajectory,Xu-multi-trajectory,Ip-LSTM-trajectory}. As the CA application trigger conditions are mobility based this appeared a promising candidate for an ML based approach. Initial implementations looked at the performance of LSTM based models as well as Gated Recurrent Units (GRU):

Both LSTMs and GRUs are recurrent neural networks that operate by maintaining a form of long term memory throughout the network. This works by persisting information across each step of the network. Through experimentation it was determined that a GRU based approach was preferable to LSTM in terms of accuracy and speed. The architecture of the GRU model is shown in Fig. \ref{fig:GRU-Arch}. The model operates by persisting a form of long term memory across a series of data. The architecture of a GRU neuron can be broken down into several components, which are highlighted in Fig. \ref{fig:GRU-Arch}. The first part being a reset gate which decides whether the previous neuron's state information \textit{ht-1} in combination with the current input data \textit{xt} will be used in the prediction. This is followed by an update gate that determines the information to be stored in the hidden state which represents the additional historical data stored by the model and used in subsequent prediction. Finally, there is the hidden state candidate which determines the \textit{ht} or the ultimate hidden state of the neuron. As illustrated in Fig. \ref{fig:GRU-rolled}, each neuron feeds into subsequent neurons i.e. taking a specific segment of the data sequence and their predictions feeding forward to later elements in the sequence. An example of this is where the data \textit{xt-1} and prediction \textit{ht-1} influences the prediction of the subsequent data \textit{xt}. This enables the form of long term memory which allows these RNN models to deal effectively with sequence data.

To apply the GRU for CA prediction, several approaches were initially employed; Initial iterations of the model predicted the exact RRI based on the 4 ETSI triggering conditions:

\begin{enumerate}
    \item Position change of $>$ 4m.
    \item Heading change of $>$ 4\degree.
    \item Speed change of $>$ 4m/s.
    \item Time since last transmission exceeds 1s.
\end{enumerate}

This proved to be insufficiently accurate, so the final solution implements 3 models which are used to predict the heading (sine and cosine) and speed of the vehicle for a 1 second interval but in 100ms stpdf (10 predictions made). The predictions are made by passing in the speed, sine and cosine, sampled every 100ms, for the last minute. The predictions are then compared to the ETSI trigger conditions and used to determine the predicted RRI. It was necessary to consider sine and cosine to eliminate inaccurate predictions e.g. where $355\degree$ is nearer to $5\degree$ than $30\degree$. Considering the cosine and sine of the heading allows for the reconstruction of the heading but avoids issues in the training process. The packet will then be passed to the lower layers with the predicted RRI so the SCI can be updated with the new RRI. 

\subsection{CA GRU Prediction Model Implementation}
\label{sec:machineLearningApplied}

As the accuracy of any proposed model will be highly dependant on having realistic training data, we employed several different data sets as shown in Table \ref{tab:scenarios}. These scenarios were used to model vehicular dynamics incorporating the CA triggering conditions to train the model effectively. The data was split into sequences of 1 minute and 1 second length with speed and heading data sampled every 100ms. The first minute represents the data to use for prediction with the final 1 second representing the ultimate prediction. Each sequence was then randomly assigned to the training, validation and test sets with a split of 80\%:10\%:10\%.

\begin{table}[htbp]
\caption{Vehicular Scenarios employed for training data generation.}
\label{tab:scenarios}
\begin{center}
\begin{tabular}{p{2cm} | l | p{4cm}}
\hline\hline 
\textbf{Scenario} & \textbf{Type} & \textbf{Description} \\
\hline
Digital Twin Geneva Motorway (DT-GM) & Highway &  Digital twin of Geneva highways allowing direct link to real time data, as well as default scenarios representing a typical day. \\
\hline
Dublin (ITSC2020 CAV impact) & Highway/City & Scenario is separated into segments of the M50 motorway around Dublin and internal segments of Dublin city itself. \\
\hline
Ingolstadt (InTas) & City & City scenario based on typical traffic on a day in Ingolstadt. \\
\hline
Monaco SUMO Traffic (MoST) & Highway/City & Detailed scenario with traffic throughout the day in Monaco including surrounding highways. \\
\hline
Turin SUMO Traffic (TuST) & Highway/City & Similar to MoST scenario is 24 hours snapshot of Turin including highways surrounding the city. \\
\hline
Cork South Ring Road & Highway & Custom implemented scenario with a 21km scenario with high speeds.\\
\hline\hline
\end{tabular}
\end{center}
\end{table}

The implementation details of the GRU prediction model is shown in Table \ref{tab:models}. The structure and parameters used for each of the layers, such as the choice between GRU and LSTM, the number of Recurrent Layers and dense layers, learning rate, neurons per layer and activation functions used was determined using a bayesian search grid on all parameters, from which the best performing model was determined and trained on the full data set. These results show good performance for the training of the model with low error across training, validation and test data. This demonstrates that the model generalises well to unseen data as well as avoiding overfitting.

\begin{table*}[t]
\caption{Structure of GRU Models and performance across data sets (MSE)}
\label{tab:models}
\begin{center}
\renewcommand{\arraystretch}{1.2}
\begin{tabular}{l l r || l l r || l l r}
\hline\hline 
\multicolumn{3}{c}{\textbf{Speed Model}} || & \multicolumn{3}{c}{\textbf{Sine Model}}  || & \multicolumn{3}{c}{\textbf{Cosine Model}}\\ [0.5ex] 
\hline
\textit{Layer} & \textit{\makecell{ Activation \\ Function }} & \textit{Neurons} & \textit{Layer} & \textit{\makecell{ Activation \\ Function }} & \textit{Neurons} &\textit{Layer} & \textit{\makecell{ Activation \\ Function }} & \textit{Neurons} \\
\hline
GRU & Tanh & 128 & GRU & Tanh & 128 & GRU & Tanh & 128 \\
GRU & Tanh & 128 & GRU & Tanh & 128 & GRU & Tanh & 128 \\
Dense & Linear & 10 & GRU & Tanh & 128 & GRU & Tanh & 128 \\
& & & Dense & Linear & 10 & Dense & Relu & 32 \\
& & & & & & Dense & Relu & 32 \\
& & & & & & Dense & Linear & 10 \\
\hline
\textit{Training} & \textit{Validation} & \textit{Test} & \textit{Training} & \textit{Validation} & \textit{Test} & \textit{Training} & \textit{Validation} & \textit{Test} \\
\hline
0.098 & 0.01 & 0.001 & 0.0147 & 0.015 & 0.0094 & 0.0018 & 0.0015 & 0.008\\
\hline\hline
\end{tabular}
\end{center}
\end{table*}

\section{Experimental Results}
\label{sec:results}

\subsection{Simulation Modelling and Scenarios.}
\label{subsec:Config&scenarios}

This section evaluates the performance of the 
GRU CA prediction model and compares it against 3 different models. Firstly, a periodic model that is configured to exhibit the same average packet inter-arrival rate as the ETSI triggered traffic. This represents the best case performance of SB-SPS, which the ML model should tend towards when it is operating most effectively. We use this to benchmark the accuracy and usefulness of our proposed approach and it is labelled as \textit{Periodic} throughout our analysis. The second model is the default operation of SB-SPS (No GB) where if no packet is available for transmission the grant is maintained but a missed transmission occurs. This is labelled \textit{Default}. The final model is where the RRI is configured to be the average packet inter-arrival rate (rounded to the closest RRI). This is labelled as \textit{Mean IPT} in subsequent discussions.  

All simulations are conducted using OMNeT++ OpenCV2X \cite{openCV2XFirst}, an open source implementation of the C-V2X standard that has been updated for the NR-V2X dynamic grant mechanism. Key simulation parameters are summarised in Table \ref{tab:cv2x-setup}.   

\begin{figure}[htbp]
\begin{subfigure}{.48\textwidth}
  \centering
  \includegraphics[width=.8\linewidth]{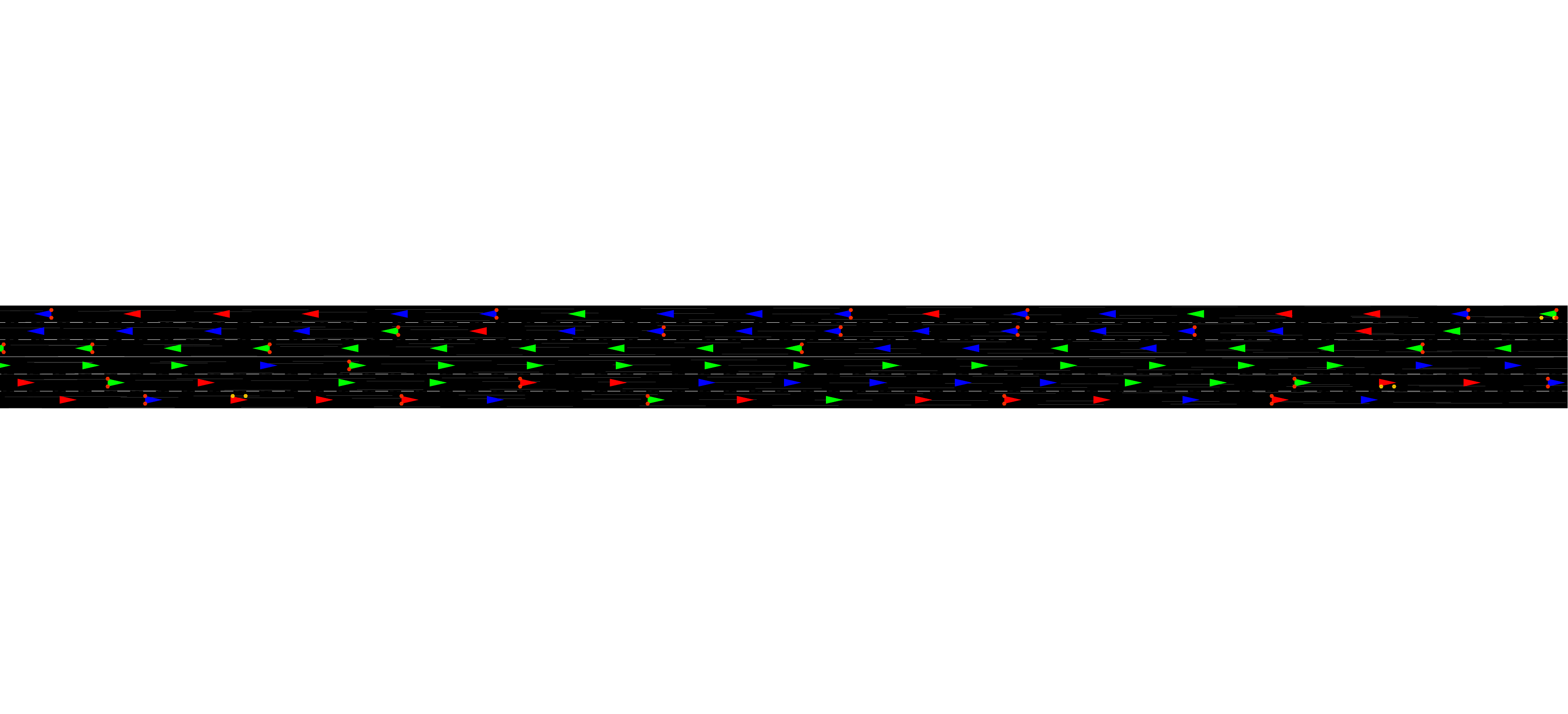}
  \caption{\textbf{Straight Highway}. Trigger conditions are positional changes only.}
  \label{fig:highway-straight-diagram}
\end{subfigure}
\begin{subfigure}{.48\textwidth}
  \centering
  \includegraphics[width=.8\linewidth]{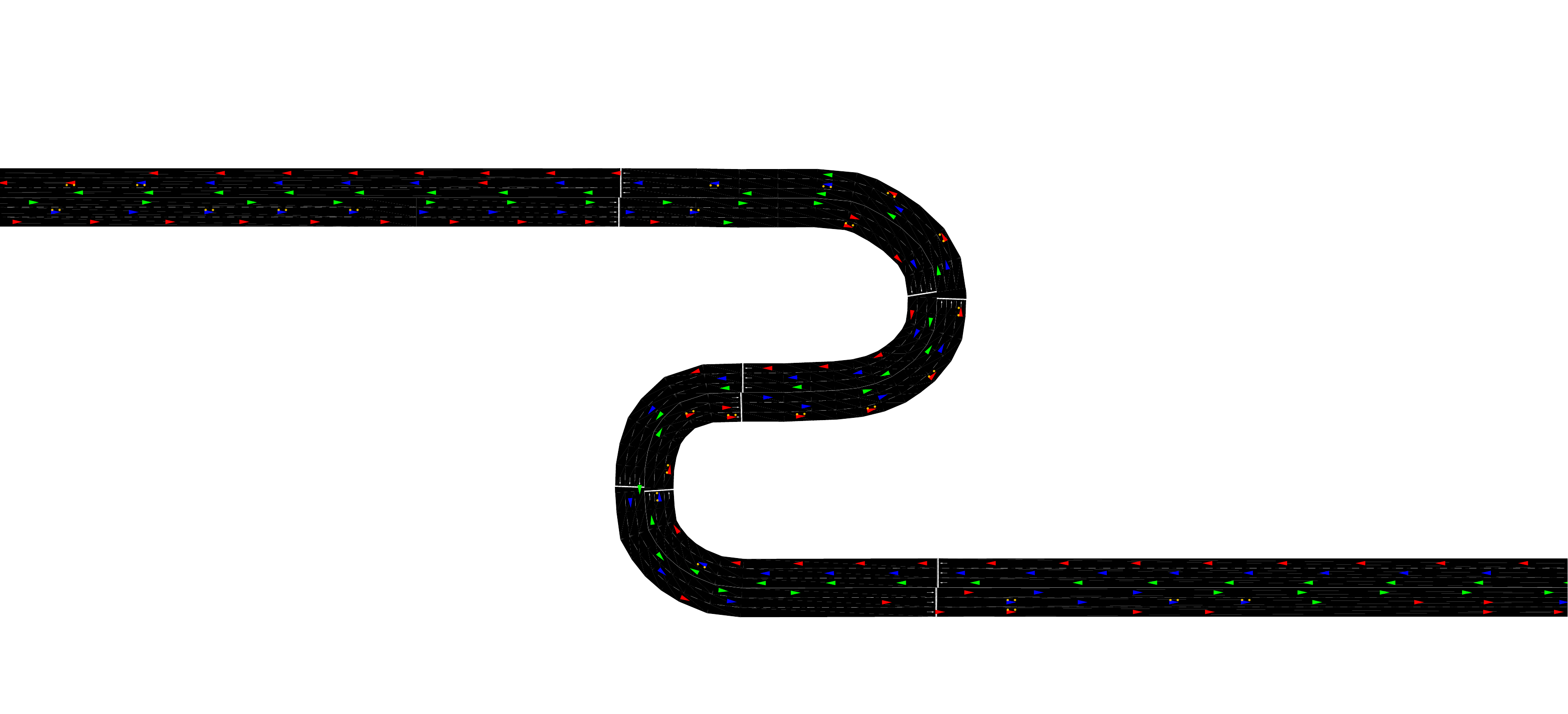}
  \caption{\textbf{Curved Highway}. Trigger conditions are positional changes or change in heading.}
  \label{fig:highway-curve-diagram}
\end{subfigure}
\begin{subfigure}{.48\textwidth}
  \centering
  \includegraphics[width=.8\linewidth]{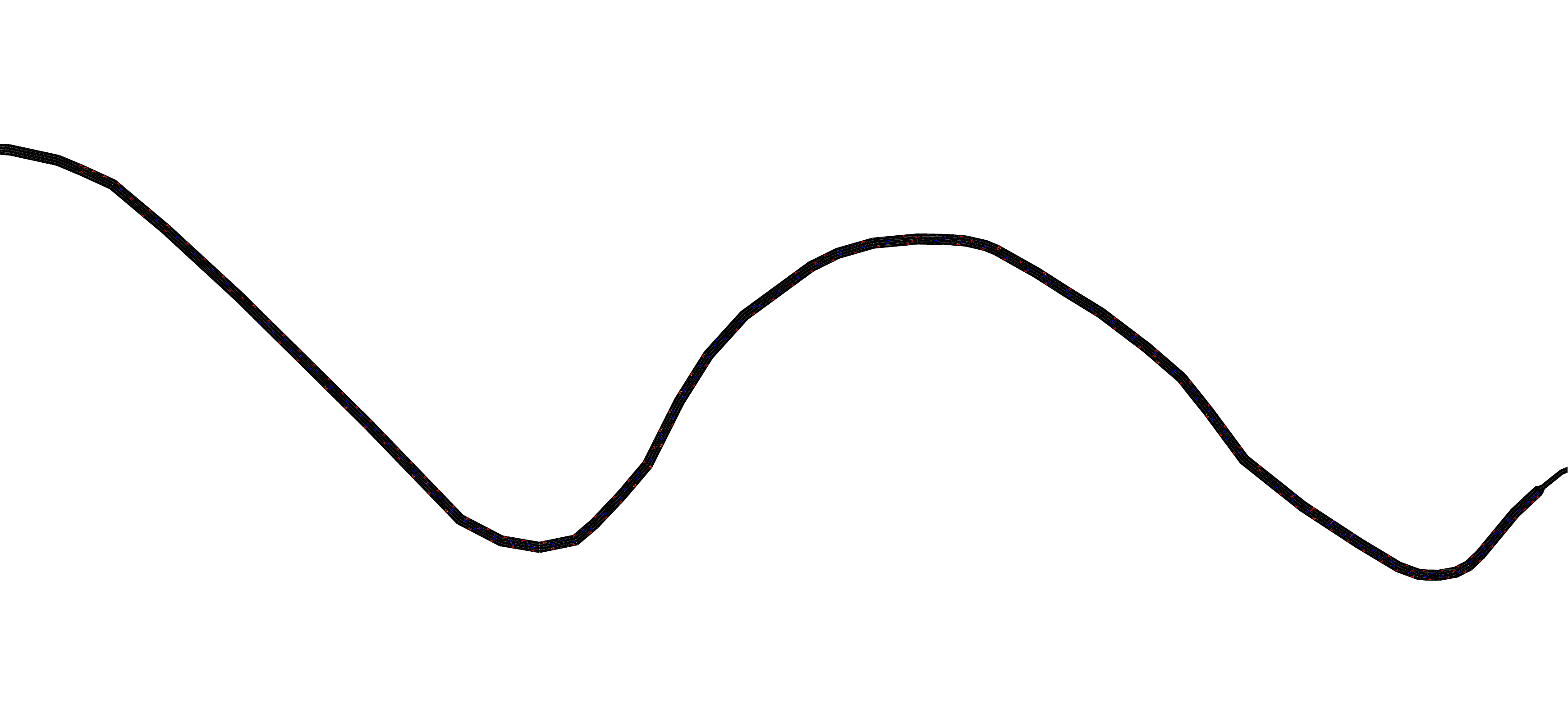}
  \caption{\textbf{Real road network segment}.}
  \label{fig:corkRoad-diagram}
\end{subfigure}
\caption{Simulated road network topologies.}
\label{fig:scenarios}
\end{figure}

\begin{table}[htbp]
\caption{Simulation Parameters.}
\begin{center}
\begin{tabular}{l l}
\hline\hline
\textbf{Parameter} & \textbf{Value}\\
\hline
\multicolumn{2}{c}{\textbf{Channel settings}}\\
\hline
Carrier frequency & 5.9 GHz\\
Channel bandwidth & 10 MHz\\
No. subchannels \& Size & 3 \& 16 RBs\\
\hline
\multicolumn{2}{c}{\textbf{Application \& Mobility}}\\
\hline
Packet size & 190 Bytes\\
Vehicle Mobility & SUMO (step-length = 10ms)\\
\hline
\multicolumn{2}{c}{\textbf{MAC \& PHY layer}}\\
\hline
Resource keep probability & 0\\
RSRP threshold & -126 dBm \\
Propagation model & Winner+ B1\\
MCS & 6 (QPSK 0.5)\\
Transmission power & 23 dBm\\
Noise figure & 9 dB\\
Shadowing variance & 3 dB\\
\hline\hline
\end{tabular}
\label{tab:cv2x-setup}
\end{center}
\end{table}

To determine the performance of the CA prediction model across diverse vehicular scenarios, three scenarios are considered. The first is a simple highway scenario as shown in Fig. \ref{fig:highway-straight-diagram}, where the trigger conditions occur exclusively due to position changes as vehicles have fixed speed. The second scenario is a curved highway as shown in Fig. \ref{fig:highway-curve-diagram} where triggering conditions occur due to both positional and heading changes. The final scenario is based on a real road network in Ireland that does not occur in the data sets. This scenario is included to show how the model can generalise to other real world environments and is shown in Fig \ref{fig:corkRoad-diagram}. 


\subsection{Performance in a straight highway (fixed speeds).}
\label{subsec:highwaystraight}

\begin{figure}[htbp]
  \centering
  \includegraphics[width=.88\linewidth]{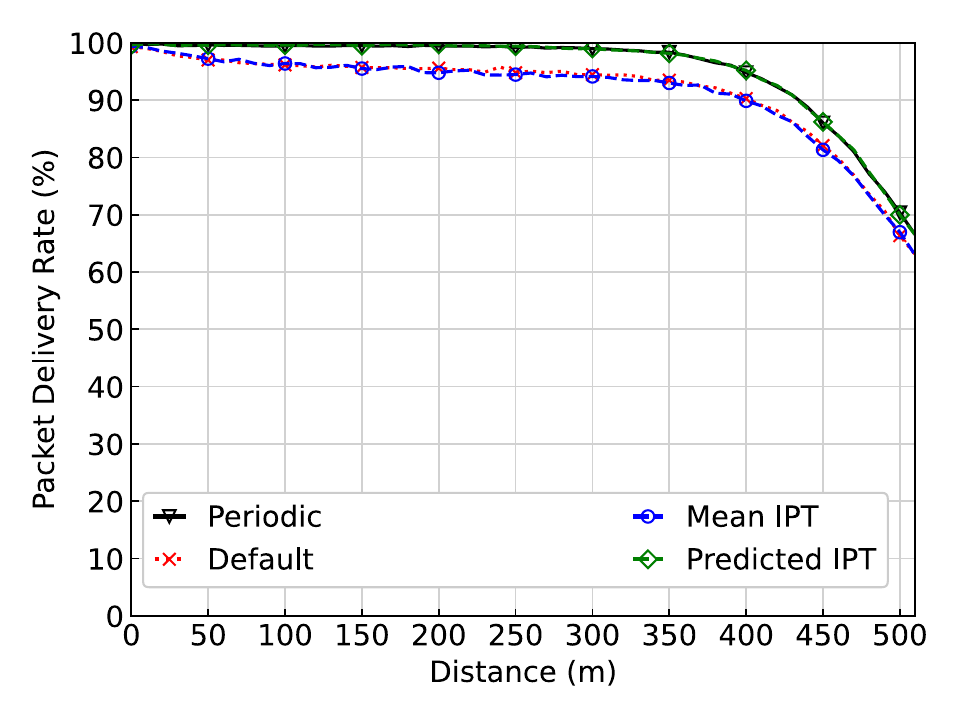}
\caption{Performance of ETSI application model in a highway scenario (fixed vehicle speed).}
\label{fig:highway-straight}
\end{figure}

\begin{table}[htbp]
\caption{Application \& RRI Performance (Highway)}
\begin{center}
\begin{tabular}{l r r r}
\hline\hline
\textbf{Scenario} & \textbf{CAM rate (ms)} & \textbf{IPT (ms)} & \textbf{$\delta$Col}\\
\hline
Periodic & 301 $\pm$ 14 & 300 $\pm$ 16 & 99\\
SB-SPS (No-GB) & 300 $\pm$ 15 & 300 $\pm$ 25 & 298\\
Mean IPT & 301 $\pm$ 21 & 302 $\pm$ 32 & 321\\
Predicted IPT & 300 $\pm$ 13 & 300 $\pm$ 15 & 98\\
\hline\\
\hline
\textbf{Scenario} & \makecell{\textbf{RRI}\\\textbf{Error (ms)}} & \textbf{Predictions} & \textbf{Inaccuracies}\\
\hline
Mean IPT & 11.613 $\pm$ 81.244 & 3308 & 163\\
Predicted IPT & 0.172 $\pm$ 5.426 & 3184 & 3\\
\hline\\
\hline
\makecell{\textbf{Prediction} \textbf{Error}} & & \textbf{Heading $\degree$} & \textbf{Abs. Speed (m/s)}\\
\hline
 & & 0.2 $\pm$ 0.2 & 0.2 $\pm$ 0.1\\
\hline\hline
\end{tabular}
\label{tab:highway-app}
\end{center}
\end{table}

The highway scenario shows the highest level of performance for the proposed solution \textit{Predicted IPT} as shown in Fig. \ref{fig:highway-straight}. It can be clearly seen that the predicted performance mimics the performance in a \textit{Periodic} configuration and out-performs the \textit{Default} and \textit{Mean IPT} setups by 5\%. Table \ref{tab:highway-app} summarises important performance aspects ranging from application and transmission rates, to RRI prediction performance and the underlying trajectory predictions. All configurations perform identically in terms of CAM rate (trigger rate) and IPT. The $\delta$Col field explains the performance differences, representing the number of transmissions experiencing interference from neighbouring vehicles (within 500m). It can be seen that \textit{Periodic} and \textit{Predicted IPT} have significantly fewer $\delta$Cols than the \textit{Default} scenario, with \textit{Mean IPT} resulting in slightly increased collisions. \textit{Predicted IPT} performs so well due to its ability to accurately predict the RRI, thus maintaining the SB-SPS grant mechanism more effectively. The \textit{Predicted IPT} only exhibits 3 errors in prediction in the highway scenario. This is as a result of highly accurate trajectory predictions with an error of only 0.2 m/s and degrees in heading.

\subsection{Performance in a curved highway (varying headings).}
\label{subsec:curved}

\begin{figure}[htbp]
  \centering
  \includegraphics[width=.85\linewidth]{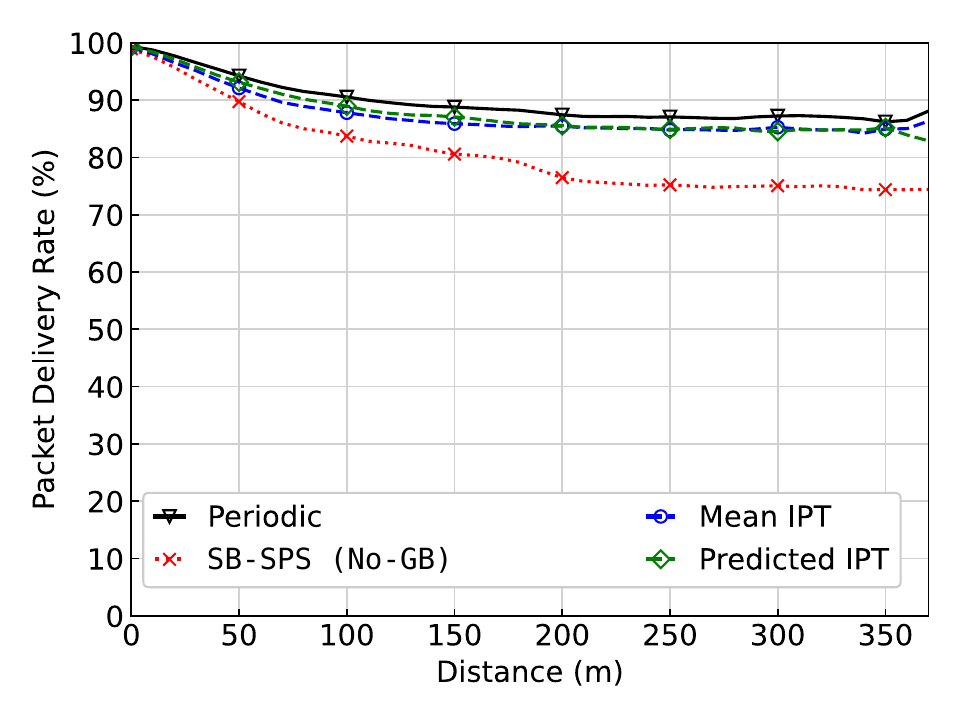}
\caption{Performance of ETSI application model in a curved highway (varying vehicle headings).}
\label{fig:curve-pdr}
\end{figure}

\begin{table}[htbp]
\caption{Application \& RRI Performance (Curved)}
\begin{center}
\begin{tabular}{l r r r}
\hline\hline
\textbf{Scenario} & \textbf{CAM rate (ms)} & \textbf{IPT (ms)} & \textbf{$\delta$Col}\\
\hline
Periodic & 266 $\pm$ 50 & 266 $\pm$ 51 & 1862 \\
SB-SPS (No-GB) & 299 $\pm$ 130 & 300 $\pm$ 131 & 2551 \\
Mean IPT & 349 $\pm$ 110 & 363 $\pm$ 100 & 1920 \\
Predicted IPT & 350 $\pm$ 111 & 359 $\pm$ 107 & 1827 \\
\hline\\
\hline
\textbf{Scenario} & \makecell{\textbf{RRI}\\\textbf{Error (ms)}} & \textbf{Predictions} & \textbf{Inaccuracies}\\
\hline
Mean IPT & 28.754 $\pm$ 126.015 & 9398 & 4340\\
Predicted IPT & 22.840 $\pm$ 139.954 & 9354 & 3279\\
\hline\\
\hline
\textbf{Prediction} \textbf{Error} & & \textbf{Heading $\degree$} & \textbf{Abs. Speed (m/s)}\\
\hline
 & & 6.9 $\pm$ 8.7 & 0.3 $\pm$ 0.4\\
\hline\hline
\end{tabular}
\label{tab:curve-app}
\end{center}
\end{table}

The performance for the second scenario (curved highway) is shown in Fig. \ref{fig:curve-pdr}

It can be observed that the PDR for both \textit{Mean-IPT} and \textit{Predicted-IPT} exhibit similar performance which is close to that of \textit{Periodic} while significantly outperforming \textit{Default}. In Table \ref{tab:curve-app}, it can be seen that \textit{Periodic} exhibits higher transmission rates while maintaining good performance. \textit{Default} also exhibits higher transmission rates which contributes to its poorer performance, highlighted further in its increased $\delta$Col value. The \textit{Mean} and \textit{Predicted IPT} setups exhibit similar performance  except where \textit{Predicted} incurs a lower number of collisions. This is not reflected in improved PDR due to the low density nature of the scenario.
This is as a result of it exhibiting fewer errors in predicting the RRI, both show similar accuracy in their predictions of the RRI though Predicted reduces the number of inaccuracies by over 1000. It is worth observing that error in terms of speed does not change significantly from the highway scenario but heading error is increased by $6.9\degree$. This increase in heading error directly results in more inaccurate predictions reducing the performance of the Predicted IPT configuration.

\subsection{Performance in an unseen real road network segment (varying speed \& heading).}
\label{subsec:unseenRoad}

\begin{figure}[htbp]
\begin{subfigure}{.48\textwidth}
  \centering
  \includegraphics[width=.88\linewidth]{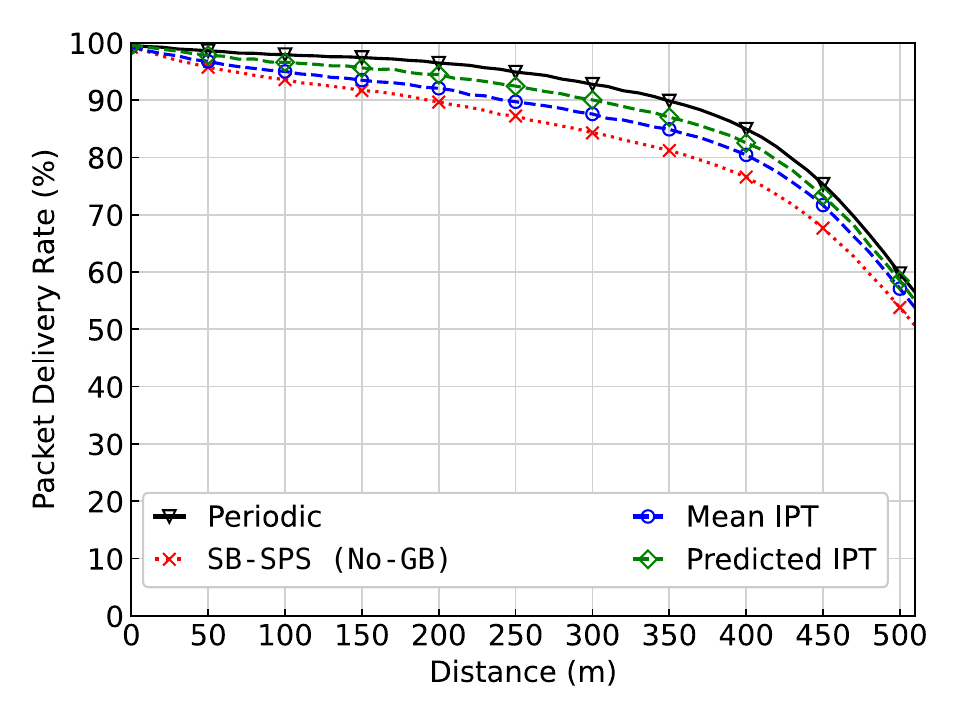}
  \caption{PDR performance}
  \label{fig:cork-pdr}
\end{subfigure}
\begin{subfigure}{.48\textwidth}
  \centering
  \includegraphics[width=.88\linewidth]{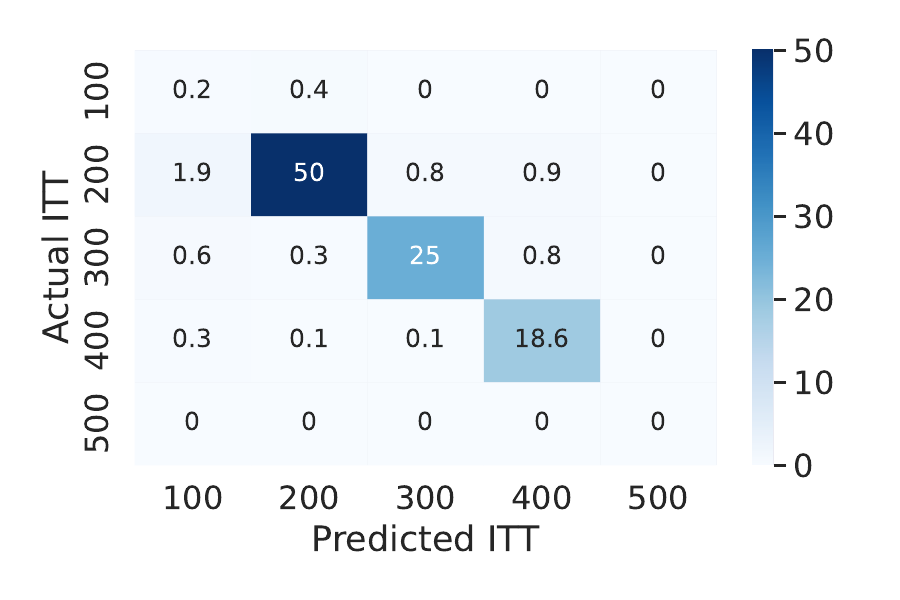}
  \caption{RRI confusion matrix (overall percentage).}
  \label{fig:cork-rri-percent}
\end{subfigure}
\caption{Performance of ETSI application model in unseen real road networking segment (varying speed and heading).}
\label{fig:cork}
\end{figure}

\begin{table}[htbp]
\caption{Application \& RRI Performance (Cork)}
\begin{center}
\begin{tabular}{l r r r}
\hline\hline
\textbf{Scenario} & \textbf{CAM rate (ms)} & \textbf{IPT (ms)} & \textbf{$\delta$Col}\\
\hline
Periodic & 213 $\pm$ 36 & 213 $\pm$ 37 & 813 \\
SB-SPS (No-GB) & 238 $\pm$ 61 & 244 $\pm$ 64 & 1836 \\
Mean IPT & 227 $\pm$ 67 & 227 $\pm$ 68 & 1441 \\
Predicted IPT & 239 $\pm$ 60 & 242 $\pm$ 59 & 1089\\
\hline\\
\hline
\textbf{Scenario} & \makecell{\textbf{RRI}\\\textbf{Error (ms)}} & \textbf{Predictions} & \textbf{Inaccuracies}\\
\hline
Mean IPT & 9.962 $\pm$ 58.945 & 18749 & 2463\\
Predicted IPT & 2.723 $\pm$ 51.731 & 18988 & 2285\\
\hline\\
\hline
\textbf{Prediction Error} & & \textbf{Heading $\degree$} & \textbf{Abs. Speed (m/s)}\\
\hline
 & & 3.1 $\pm$ 5.0 & 0.3 $\pm$ 0.4\\
\hline\hline
\end{tabular}
\label{tab:cork-app}
\end{center}
\end{table}

As can be observed in Fig. \ref{fig:cork-pdr}, the \textit{Predicted IPT} results demonstrates improved PDR over \textit{Default} and also outperforms \textit{Mean IPT}. As shown in Table \ref{tab:cork-app}, there are some differences in terms of IPT for the configurations with \textit{Periodic} exhibiting higher transmission rates, along with \textit{Mean IPT}. Ultimately, it can be seen that \textit{Predicted IPT} significantly reduces the number of collisions and closes in on the \textit{Periodic} performance. In this scenario, we again observe \textit{Predicted IPT} exhibiting lower RRI error and reducing the number of inaccurate predictions, which explains the performance improvement when compared with \textit{Mean IPT}. When analysing the trajectory predictions it is clear that speed predictions are highly accurate and heading error is observed at $3.1\degree$. How this impacts the performance is further depicted in Fig. \ref{fig:cork-rri-percent}, which shows a confusion matrix for predicted IPT versus the actual IPT as a percentage of all predictions. Most of the IPTs are concentrated around 200ms and 300ms. Overall the model makes accurate predictions 88.2\% of the time. The primary cause of inaccuracy is due to heading errors in prediction, either under-estimating the heading change (100ms IPT predicted as 200ms) or over-estimating it (200ms IPT predicted as 100ms). This was highlighted as the error in heading can exceed the $4\degree$ threshold for CAM triggering. 



\section{Conclusion}
\label{sec:conclusion}

This paper introduces a machine learning (GRU) based approach to accurately predict the inter-packet times of CAMs, based on the current vehicle dynamics. This prediction enables the improvement of the SB-SPS scheduler when dealing with aperiodic application traffic patterns by adaptively updating the RRI field in SCI packets. We have shown that this model is able to accurately predict and update the RRI in a diverse set of scenarios, including an unseen real world road network thereby improving the scheduling performance of SB-SPS. This approach can be utilised for other similar vehicular applications such as Cooperative Perception, and should enable SB-SPS to operate effectively for core vehicular services that exhibit  aperiodic characteristics. 

\section*{Acknowledgements}
This publication has emanated from research conducted with the financial support of Science Foundation Ireland (SFI) under Grant No: 17/RC-PhD/3479. For the purpose of Open Access, the author has applied a CC-BY public copyright licence to any Author Accepted Manuscript version arising from this submission.

\bibliographystyle{IEEEtran}
\bibliography{references}

\end{document}